\documentclass[12pt]{article}
\usepackage{sc3conf}
\usepackage{amsfonts}
\usepackage{epsfig}

\begin{document}
\raggedbottom

\title{
BPS Walls and Junctions 
 in ${\cal N}=1$ SUSY % Supersymmetric 
Nonlinear Sigma Models 
}

\authors{Masashi~Naganuma,\adref{1}
  Muneto~Nitta,\adref{2}  
  and Norisuke~Sakai\adref{1,3}}

\addresses{\1ad Department of Physics, Tokyo Institute of 
Technology, %\\
Tokyo 152-8551, JAPAN 
  \nextaddress \2ad Department of Physics, Purdue University, 
West Lafayette, IN 47907-1396, USA
  \nextaddress \3ad Speaker at the conference.
}

\maketitle

\begin{abstract}
%  In the \LaTeX\ source file, the Abstract should go \textit{after}
%  the \verb"\maketitle" command.
The ${\cal N}=1$ 
SUSY nonlinear sigma models in four spacetime dimensions 
are studied to obtain BPS walls and junctions. 
A nonlinear sigma model with a single chiral scalar superfield 
is found which has a moduli space of the topology of $S^1$ 
and admits BPS walls and junctions connecting arbitrary points in moduli 
space. 
New BPS junction solutions connecting $N$ discrete vacua 
are also found 
for nonlinear sigma models with several chiral scalar superfields. 
More detailed exposition of our results can be found in Ref.\cite{NNS}. 
\end{abstract}

\section{Introduction}
Supersymmetric theories provides the most attractive possibility 
for unified theories beyond the standard model. 
 \cite{DGSWR}. 
In recentl years there have been intensive studies on 
models with extra dimensions, \cite{LED}, \cite{RS} where 
our world is assumed 
to be realized on an extended topological defects 
such as domain walls or various branes. 
Supersymmetry can be combined with this brane-world scenario 
and helps the construction of the extended objects. 

In order to consider models with extra dimensions, 
we need to discuss 
supersymmetric theories in spacetime with dimensions higher than four. 
They should have at least eight supercharges. 
These SUSY are so restirictive that 
the nontrivial interactions require the nonlinearity of 
kinetic term (nonlinear sigma model) if there are only hypermultiplets 
 \cite{AlvarezFreedman}.

Our purpose is to study the nonlinear sigma model 
in a simpler context of ${\cal N}=1$ SUSY theories in four dimenions 
to obtain walls and/or junctions as BPS solutions. 
This is interesting in itself, and will serve as a starting 
point for a more difficult case of larger number of SUSY charges. 
We find a nonlinear sigma model with a single chiral 
scalar field is also obtained which admits our $Z_3$ symmetric 
junction solution as an exact BPS solution. 
We find that this single field model has a moduli space with $S^1$ topology 
and that it admits BPS walls 
and junctions connecting arbitrary points in the moduli space. 
We also find other nonlinear sigma models with several chiral 
scalar superfields 
which admit a new exact junction solutions connecting $N$ discrete 
vacua.  
The model and the solution are generalizations of our original 
 $Z_3$ symmetric vacua to a generic $N$ discrete vacua including models 
 with arbitrary number of asymmetric vacua.  
We deal with classical field theories in this paper and 
will postpone to discuss questions on quantization and quantum 
effects for subsequent studies.

\section{Nonlinear sigma models with exact junction 
solutions}
\label{sc:NLSM_junction}
Requiring a single conserved  supercharge for a general nonlinear sigma model 
with a K\"ahler potential $K$ and a superpotential ${\cal W}$, the $1/4$ BPS 
equation is given by \cite{AbrahamTownsend} 
\begin{equation}
2 
{\partial A^i \over \partial z} 
= \Omega \; K^{i j^*} \! (A, A^*) \; 
\frac{\partial {\cal W}^*}{ \partial A^{*j}}, 
\label{Be2}
\end{equation}
where $z=x^1+i x^2$ is the complex 
coordinate, $K^{ i j^*}$ is the K\"ahler metric of the nonlinear sigma model, 
 and the constant phase factor $\Omega$ is given by the 
central charges which correspond to boundary conditions 
determining the orientation of junction.  
Previously we have obtained the first analytic solution of $1/4$ BPS 
junction in a model with a number of chiral scalar fields 
with $U(1)\times U(1)$ gauge interactions \cite{OINS}. 
The same junction configuration turns out to be 
a BPS solution of a simpler model with only 
chiral scalar fields. 

There are four chiral scalar superfields ${\cal M}_i, i=1,2,3$ and $\hat T$ 
with the 
minimal kinetic term. 
The linear sigma model with $Z_3$ symmetric junction 
solution can be 
reduced to the following form of superpotential 
\begin{equation}
{\cal W}= {1 \over 2}\left[ \sqrt{3} \hat T - 
\sum_{j=1}^3 \left(\sqrt{3}\hat T-{\rm e}^{i{2\pi \over 3}j}\right) 
{\cal M}_j^2 \right]. 
\label{eq:Z3_superpotential}
\end{equation}

There are three isolated SUSY vacua at 
$%\begin{equation}
\hat T={1 \over \sqrt{3}}{\rm e}^{i{2\pi \over 3}j}, 
\quad {\cal M}_j=\pm 1, \quad {\cal 
M}_k=0, \; 
k\not=j.
$%\end{equation}
(j=1, 2, 3). 
%as shown in Fig.\ref{Z3_linear_model} (a). 
%\begin{figure}[htbp]
%\begin{center}
%%%%%%%%%%%%%%%%epsfig.sty%%%%%%%%%%%
%\leavevmode
%\begin{eqnarray*}
%\begin{array}{cc}
%  \epsfxsize=3.5cm
%  \epsfysize=3.5cm
%\epsfbox{vacua-6.eps} & 
%  \epsfxsize=3.5cm
%  \epsfysize=3.5cm
%\hspace*{1cm}
%\epsfbox{junct-3b.eps} 
%\\
%\mbox{\footnotesize (a) Vacua of $Z_3$ linear sigma model} & 
%\hspace*{1cm}
%\mbox{\footnotesize (b) $Z_3$ junction in the base space}
%\end{array} 
%\end{eqnarray*} 
%\caption{$Z_3$ symmetric linear sigma model.  }
%\label{Z3_linear_model}
%\end{center}
%\end{figure}
%%%%%%%%%%%%%%%%%%
The ${1 \over 4}$ BPS equations (\ref{Be2}) with $\Omega=1$ turn out 
\begin{equation}
2 
{\partial {\cal M}_j \over \partial z} 
=  \left( {\rm e}^{-i{2\pi \over 3}j} - \sqrt{3}\hat T^*\right) {\cal M}_j^*, 
\end{equation}
\begin{equation}
2 
{\partial \hat T \over \partial z} 
= {\sqrt{3} \over 2}\left( 1 - \sum_{j=1}^3{\cal M}_j^{*2}\right). 
\label{eq:T_BPS_Z3_LSM}
\end{equation}
It is convenient to define the following auxiliary 
 quantities taking real values 
\begin{equation}
f_j \equiv \exp \left({1 \over 2}
\left({\rm e}^{-i{2\pi \over 3}j}z
+{\rm e}^{i{2\pi \over 3}j}z^*\right) \right), 
 \qquad j=1, 2, 3. 
\label{eq:fj_Z3}
\end{equation}
which satisfy %the following identity \begin{equation}
$\prod_{j=1}^3 f_j =1$. %\end{equation}
The exact BPS solution for a junction connecting three 
vacua is given by \cite{OINS} 
\begin{equation}
\hat T = {1 \over \sqrt{3}}
{\sum_{j=1}^3 {\rm e}^{i{2\pi \over 3}j}f_j \over \sum_{k=1}^3 
f_k}, 
\qquad 
{\cal M}_j = 
{f_j \over \sum_{k=1}^3 f_k}. 
\label{eq:junction_Z3_LSM}
\end{equation}
We have shown that the configuration (\ref{eq:junction_Z3_LSM}) 
on a circle at infinity $|z|\rightarrow \infty$ reduces to a 
collection of three walls separating three vacuum regions, 
which are represented by three straight line segments connecting 
three vacua in the complex plane of superpotential \cite{OINS}. 

We observe that the junction solution is a mapping from a 
two-dimensional 
base space $x^1, x^2$ to the complex scalar fields ${\cal 
M}_j, j=1,2,3$ and 
$\hat T$. 
Therefore we can reexpress the complex scalar fields ${\cal 
M}_j, j=1,2,3$ 
in favor of the complex coordinate $z=x^1+ix^2$ and then 
invert the relation 
between $\hat T$ and $z$ to express everything in favor of $\hat T$ 
eventually. 
In particular, we can express the right-hand side of the BPS 
equation for $\hat T$ in Eq.(\ref{eq:T_BPS_Z3_LSM}) 
solely in terms of $\hat T$ by this process. 
The resulting BPS equation can be interpreted as the BPS 
equation 
(\ref{Be2}) 
in a nonlinear sigma model with appropriate 
superpotential and 
K\"ahler metric. 
Using Eqs.(\ref{eq:T_BPS_Z3_LSM}), 
and (\ref{eq:junction_Z3_LSM}), we obtain 
\begin{eqnarray}
2 
{\partial \hat T \over \partial z} 
&\!\!\!
=
&\!\!\!
{\sqrt{3} \over 2}\left(1 - \sum_{j=1}^3{\cal M}_j^{*2} \right)
= {\sqrt{3} \over 2}\left( 1 - 
{\sum_{j=1}^3 f_j^2 \over \left(\sum_{k=1}^3 f_k\right)^2} \right)
= \sqrt{3}{f_1 f_2 + f_2 f_3 + f_3 f_1 \over \left(\sum_{k=1}^3 
f_k\right)^2}  
\nonumber \\
&\!\!\!
=
&\!\!\!
 \sqrt{3}{f_1 f_2 + {1 \over f_1} + {1 \over f_2} \over 
\left(f_1 + f_2 +{1 \over f_1 f_2}\right)^2}  
=
 \sqrt{3}{f_1 f_2\left(f_1^2 f_2^2 +  f_1 + f_2\right) \over 
\left(f_1 f_2\left(f_1 + f_2\right) +1 \right)^2}, 
\end{eqnarray}
where we have eliminated $f_3$ by means of the identity 
$\prod_j f_j=1$. 
Similarly we can reexpress the field $\hat T=\hat T_R +i \hat T_I$ as 
\begin{equation}
\sqrt{3} \hat T_R 
= {-{f_1+ f_2 \over 2}  + f_3  
\over \sum_{k=1}^3 f_k}  
= {-{f_1 f_2(f_1+ f_2) \over 2}  + 1   
\over f_1 f_2(f_1 + f_2) +1 }  
\end{equation}
\begin{equation}
\sqrt{3} \hat T_I 
= { {\sqrt3 \over 2} (f_1- f_2) 
\over f_1 + f_2 +{1 \over f_1 f_2}}  
= { {\sqrt3 \over 2} f_1f_2(f_1- f_2) \over f_1f_2(f_1 + f_2) +1 } . 
\end{equation}
Therefore we finally find that 
\begin{equation}
2 
{\partial \hat T \over \partial z} 
= { 1 \over \sqrt{3}} \left(1 - 3 \hat T_R^2 - 3 \hat T_I^2\right) 
= { 1 \over \sqrt{3}} \left(1 - |\sqrt{3}\hat T|^2\right) .
\end{equation}
The resulting equation should be identified with a ${1 \over 4}$ BPS 
equation (\ref{Be2}) 
for a nonlinear sigma 
model. 
We find that the superpotential is linear in $\hat T$ and the K\"ahler metric is nontrivial 
\begin{equation}
K_{\hat T  \hat T^*}
=  {3 \over 1 -  |\sqrt{3}\hat T|^2 }, 
\qquad {\cal W} =\sqrt{3}\hat T .
\end{equation}
We emphasize that the above choice is not a matter of convenience, 
and that the holomorphy forces us to choose the superpotential as 
the chiral scalar superfield $\hat T$ itself except for the freedom of a possible 
proportionality constant ${\cal W}/\hat T$. 
The K\"ahler potential is given by 
\begin{equation}
K(\hat T, \hat T^*)
= \int^{3|\hat T|^2}{dx \over x}\log{1 \over 1 -  x } .
\end{equation}
{}From the above procedure, we see that the nonlinear sigma 
model with a single chiral scalar superfield 
is unique modulo usual freedom of holomorphic field redefinitions, 
if we require that it allows the exact junction 
(\ref{eq:junction_Z3_LSM}) as the ${1 \over 4}$ BPS solution. 

{}By using a rescaled field $T \equiv \sqrt{3} \hat T$, 
 we obtain the nonlinear sigma model 
\begin{equation}
{\cal L}_{{\rm NLSM}}
=
-{1 \over 1-| T|^2} \partial_\mu  T^* \partial^\mu  T 
-\left(1-| T|^2\right) .
\label{eq:junction_NLSM}
\end{equation}
We observe that the nonlinear sigma model has 
 continuous vacua at $| T|=1$, corresponding to the 
continuous 
family of singularities of the K\"ahler metric. 
The moduli space has a topology of a circle. 

\begin{figure}[thbp]
\begin{center}
%%%%%%%%%%%%%%%%epsfig.sty%%%%%%%%%%%
\leavevmode
\begin{eqnarray*}
\begin{array}{cc}
  \epsfxsize=3.5cm
  \epsfysize=3.5cm
\epsfbox{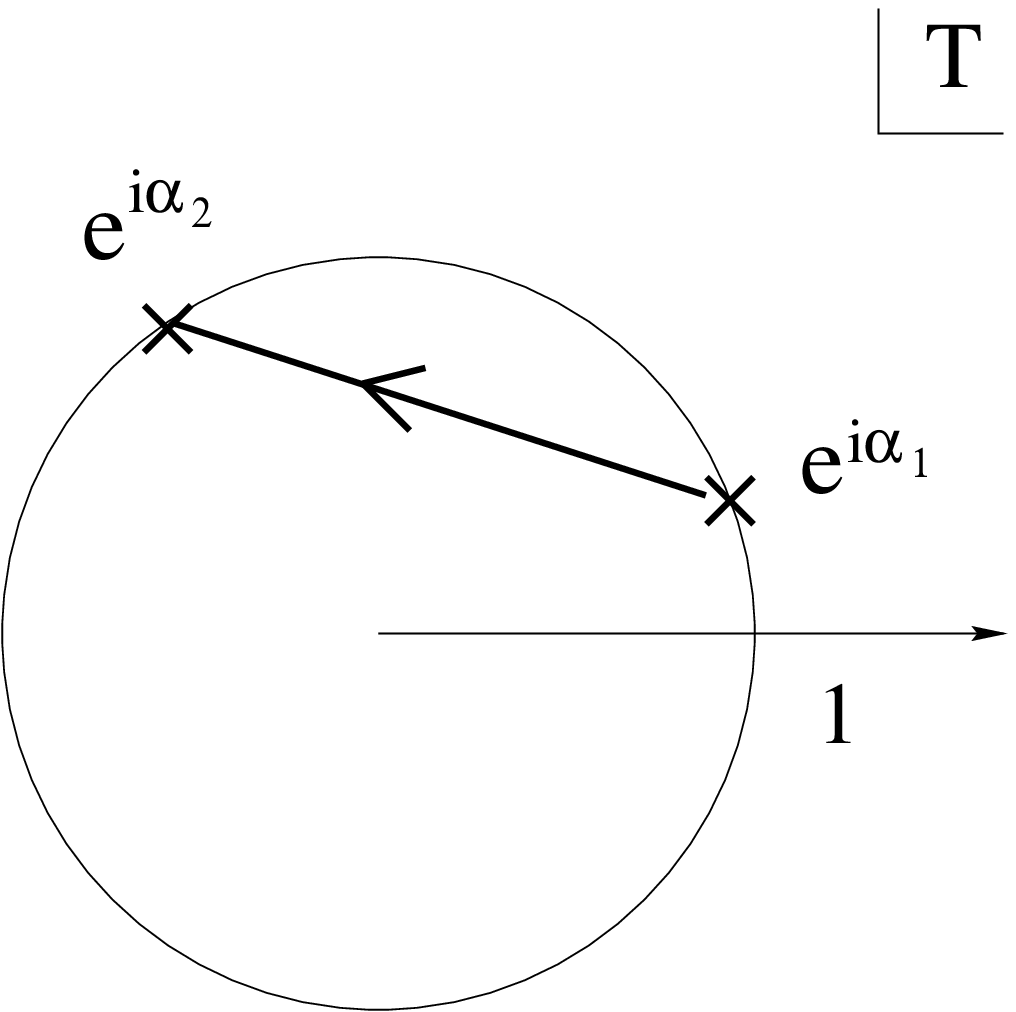} & 
  \epsfxsize=3.5cm
  \epsfysize=3.5cm
\hspace*{1cm}
\epsfbox{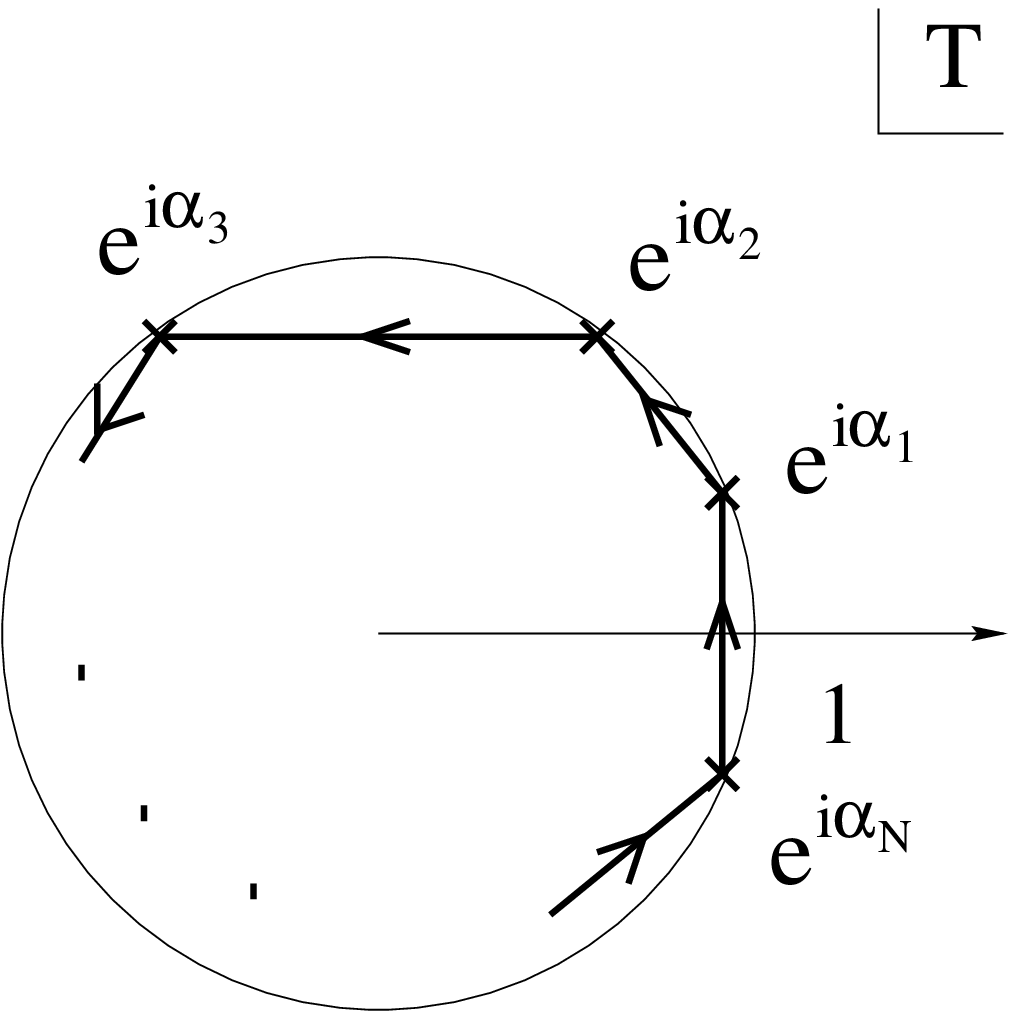} 
\\
\mbox{\footnotesize (a) A BPS wall connecting any two vacua. } & 
\hspace*{1cm}
\mbox{\footnotesize (b) Moduli space and $N$ vacua for a junction.
%$N$ junction configuration. 
}
\end{array} 
\end{eqnarray*} 
\caption{Wall and Junction in NLSM with single chiral scalar superfield. 
%(\ref{eq:junction_NLSM})
}
\label{FIG:n_junction_NLSM}
\end{center}
\end{figure}

Now we shall show that there are ${1 \over 2}$ BPS wall 
solutions 
connecting any two arbitrary points in this moduli space. 
Since an orthogonal section of the ${1 \over 2}$ BPS configuration 
should follow a straight line 
trajectory in the 
complex plane of the superpotential \cite{NNS}, we just need to 
construct a straight line 
connecting two vacua in the moduli space, thanks to our 
choice of the 
superpotential as the field variable. 
Let us define a variable taking a real value $-1 \le \tilde{ T} 
\le 1$ 
along the straight line connecting from 
$  T={\rm e}^{i \alpha_1}$ 
and 
$  T={\rm e}^{i \alpha_2}$ 
as shown in Fig. \ref{FIG:n_junction_NLSM} (a)  
\begin{equation}
  T
={{\rm e}^{i \alpha_2} + {\rm e}^{i \alpha_1} \over 2} 
+{{\rm e}^{i \alpha_2} - {\rm e}^{i \alpha_1} \over 2} \tilde{T} .
\end{equation}
By assuming the one-dimensional profile, we obtain the ${1 \over 2}$ BPS 
equation 
%(\ref{Be2}) and (\ref{Be1}) 
for the wall 
%reduce to 
\begin{equation}
{d \tilde{T} \over d x} 
= \sin {\alpha_2-\alpha_1 \over 2} 
 \left[ 1-\left(\tilde{T}\right)^2 \right] 
\end{equation}
where we have taken the constant phase factor in Eq.(\ref{Be2}) as 
$\Omega=i{\rm e}^{i {\alpha_2+\alpha_1 \over 2}}$ in order to 
orient the wall along 
the $x=x^1$ direction. 
We can recognize the familiar BPS equation to give the wall 
solution and find 
\begin{equation}
\tilde{ T} 
= \tanh 
 \left( x \sin{\alpha_2-\alpha_1 \over 2} \right)
\end{equation}
\begin{equation}
  T 
= 
{{\rm e}^{i \alpha_2}{\rm e}^{x \sin {\alpha_2-\alpha_1 \over 2}} 
+{\rm e}^{i \alpha_1}{\rm e}^{-x \sin {\alpha_2-\alpha_1 \over 2}} 
\over {\rm e}^{x \sin {\alpha_2-\alpha_1 \over 2}} +
{\rm e}^{-x \sin {\alpha_2-\alpha_1 \over 2}}} .
\label{eq:domainwall}
\end{equation}
%%%%%%%%%%%%%%%%%%
%%%%%%%%%%%%%%%%%%

More surprisingly, we can show that there exist 
exact ${1 \over 4}$ BPS junction solutions connecting any 
number of 
ordered points ${\rm e}^{i\alpha_j}, j=1, \cdots, N$ in the 
moduli space as shown in Fig.~\ref{FIG:n_junction_NLSM} (b). 
To obtain the junction solution, let us define the following 
 auxiliary quantities $f_j$ generalized from Eq.(\ref{eq:fj_Z3}) 
\begin{equation}
f_j \equiv \exp \left({1 \over 2}\left({\rm e}^{-i\alpha_j} z
+{\rm e}^{i\alpha_j}z^*\right) \right), 
\qquad j=1, \cdots, N .
\label{eq:fj_general}
\end{equation}
Let us define an auxiliary variable $\eta$ as 
$%\begin{equation}
\eta \equiv 2\log \left( \sum_{j=1}^N f_j \right) . 
%\label{eq:eta}
$%\end{equation}
Then we obtain 
\begin{equation}
{\partial \eta \over \partial z} =
 {\sum_{k=1}^N {\rm e}^{-i\alpha_k} f_k \over \sum_{j=1}^N f_j}, 
 \qquad 
{\partial^2 \eta \over \partial z \partial z^*} =
{1 \over 2} \left[1 - \left| {\partial \eta \over \partial z} 
\right|^2 
\right], 
\end{equation}

\begin{figure}[htbp]
\begin{center}
%%%%%%%%%%%%%%%epsfig.sty%%%%%%%%%%%
\leavevmode
\epsfxsize=8cm
\epsfysize=5cm
\centerline{\epsfbox{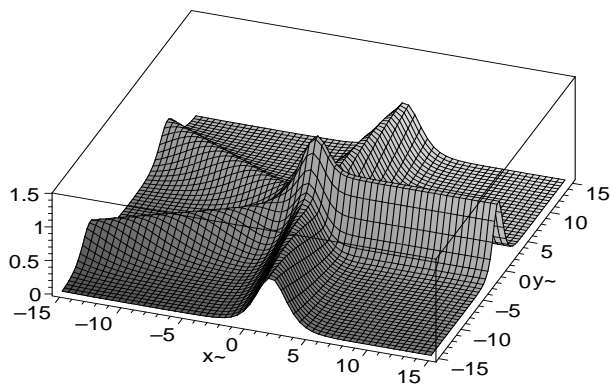}}  
\caption{Energy density of a junction in $x^1, x^2$ plane
with vacua at $\alpha_1={\pi \over 4}, \alpha_2={2\pi \over 3}, 
\alpha_3=\pi, \alpha_4={4 \pi \over 3}, \alpha_5={7\pi \over 4}$. }
\label{FIG:5_junction}
\end{center}
\end{figure}

If we take the following Ansatz 
\begin{equation}
  T = {\partial \eta \over \partial z^*} 
= {\sum_{k=1}^N {\rm e}^{i\alpha_k} f_k \over \sum_{j=1}^N f_j}, 
\label{eq:junctionAnsatz}
\end{equation}
we find that the field $T$ satisfies the ${1 \over 4}$ BPS 
equation (\ref{Be2}) 
for the nonlinear sigma model (\ref{eq:junction_NLSM}) 
with the K\"ahler metric $K_{  T  T^*}=1/(1-|  T|^2)$ and the 
superpotential ${\cal W}=T$
\begin{equation}
2{\partial  T \over \partial z} 
=1 - \left| {\partial \eta \over \partial z} \right|^2 
=1 - \left|  T \right|^2 . 
\label{eq:junction_Be_NLSM}
\end{equation}
The energy density of the junction with a choice of $5$ points in moduli 
space is shown in  Fig.~\ref{FIG:5_junction}.  

The kinetic term of the nonlinear sigma model is nonnegative 
definite 
only for $| T| \le 1$. 
It is interesting to observe that the moduli space $| T|=1$ of 
the nonlinear 
sigma model forms a natural boundary of the field space 
beyond 
which the kinetic term of the nonlinear sigma model 
is no longer positive definite. 

Another peculiar feature of the nonlinear sigma model 
(\ref{eq:junction_NLSM}) is that the boundary condition can 
be deformed 
continuously since the moduli space of vacua is continuous. 
The walls and junctions are stable as long as the boundary 
condition is fixed. 
However, the adiabatic change of the boundary condition can 
make the adiabatic 
deformation of the walls and junctions. 

\section{
 NLSM with discrete vacua and junction solution 
}
We shall work out the model which has the $N$-junction as 
a solution of ${1 \over 4}$ BPS equation. 
Let us introduce the chiral scalar superfields ${\cal M}_j, j=1, 
\cdots, N$ 
besides the chiral scalar superfield $T$. 
We assume the minimal kinetic term for these additional fields 
${\cal M}_j$. 
However, we leave the K\"ahler metric $K_{TT^*}$ of the 
field $T$ to be an arbitray function of $T, T^*$ and 
will determine it by requiring that the model possesses a 
$N$-junction as a solution of the ${1 \over 4}$ BPS 
equation. 
We assume a %$Z_N$ 
generalization of the superpotential 
in Eq.(\ref{eq:Z3_superpotential})
\begin{equation}
{\cal W}= {1 \over 2}\left[ T - 
\sum_{j=1}^N \left(T-{\rm e}^{i\alpha_j}\right) 
{\cal M}_j^2\right] 
\label{eq:N_superpotential}
\end{equation}
where the parameters $\alpha_j, \; j=1,\cdots,N$ specify the position 
of discrete vacua as we show below, and becomes 
$\alpha_j=2\pi j/N$ for the $Z_N$ symmetric case. 

The SUSY vacuum condition %(\ref{eq:SUSYcondition}) 
gives two conditions in this model 
\begin{equation}
 \left({\rm e}^{i\alpha_j}-T\right) {\cal M}_j=0, \qquad 
j=1, \cdots, N, 
\label{eq:mj_susy_cond}
\end{equation}
\begin{equation}
K^{T T^*} {1 \over 2}\left| 1 - 
\sum_{j=1}^N {\cal M}^2_j\right|^2=0. 
\label{eq:t_susy_cond}
\end{equation}
We find that there are two types of the SUSY vacua 
%as in Fig.\ref{FIG:n_junction_NLSM}(a)
: 
\begin{enumerate}
\item 
The $j$-th condition (\ref{eq:mj_susy_cond}) is satisfied 
if the field T takes a particular value $T={\rm e}^{i\alpha_j} $ 
for an integer $j$. 
Then the other conditions (\ref{eq:mj_susy_cond}) imply ${\cal M}_k=0$ 
for $k\not=j$. 
Assuming that the 
 K\"ahler metric is not singular at the point $T={\rm e}^{i\alpha_j} $, 
the second condition (\ref{eq:t_susy_cond}) is only satisfied 
by ${\cal M}_j=\pm 1$. 
\begin{equation}
T={\rm e}^{i\alpha_j}, \quad {\cal M}_j=\pm 1, \quad {\cal 
M}_k=0, \; 
k\not=j, \qquad j=1, \cdots, N. 
\label{eq:ZNvacLSM}
\end{equation}
We will check later that the K\"ahler metric is indeed not singular at this 
point. 
This is the $N$ discrete SUSY vacua given by stationary points of 
the superpotential. %as shown in Fig.~\ref{FIG:n_junction_NLSM}(a). 
\item 
If $T \not= {\rm e}^{i\alpha_j}, \; j=1, \cdots, N$, 
all the other fields have to vanish ${\cal M}_j=0$ to satisfy 
Eq.(\ref{eq:mj_susy_cond}). 
The other condition (\ref{eq:t_susy_cond}) can only be satisfied 
by a singularity of the K\"ahler metric 
\begin{equation}
K_{TT^*}=\infty. 
\label{eq:kahlersing}
\end{equation}
\end{enumerate}

The ${1 \over 4}$ BPS equations 
(\ref{Be2}) 
for ${\cal M}_j$ and $T$ are 
given by 
\begin{equation}
2 
{\partial {\cal M}_j \over \partial z} 
=  \left(  {\rm e}^{-i\alpha_j}-T^*\right) {\cal M}_j^*, 
\label{eq:BPSeq_M}
\end{equation}
\begin{equation}
2 
{\partial T \over \partial z} 
=  K^{TT^*}{1\over 2} \left( 1 - \sum_{j=1}^N{\cal M}_j^{*2}\right) .
\label{eq:T_BPS_ZN_LSM}
\end{equation}
Using the auxiliary quantities $f_j$ defined in Eq.(\ref{eq:fj_general}), 
we assume the following Ansatz for the junction 
\begin{equation}
T = 
{\sum_{j=1}^N {\rm e}^{i\alpha_j}f_j \over \sum_{k=1}^N 
f_k}, 
\qquad 
{\cal M}_j = 
{f_j \over \sum_{k=1}^N f_k}. 
\label{eq:junction_ZN_LSM}
\end{equation}
By using identities
\begin{equation}
{\partial f_j \over \partial z} = 
{1 \over 2} {\rm e}^{-i\alpha_j}f_j, 
\end{equation}
%\qquad 
\begin{eqnarray}
{\partial \over \partial z} \left({ f_j \over \sum_{k=1}^N f_k}\right)
&\!\!\!
=
&\!\!\!
%{1 \over 2} {{\rm e}^{-i\alpha_j} f_j \over \sum_{k=1}^N f_k}
%-{f_j  \over 2} 
%{\sum_m {\rm e}^{-i\alpha_m}f_m \over \left(\sum_{k=1}^N f_k\right)^2}
%=
 { f_j \over \sum_{k=1}^N f_k}
{1 \over 2} \left( {\rm e}^{-i\alpha_j}-T^* \right)  
\end{eqnarray}
we find that the ${1 \over 4}$ BPS equation for ${\cal M}_j$ 
in Eq.(\ref{eq:BPSeq_M}) is satisfied. 
Then the remaining ${1 \over 4}$ BPS equation 
for $T$ is also satisfied if and only if the 
 K\"ahler metric of the field $T$ is given by 
\begin{eqnarray}
K_{TT^*} 
&\!\!\!
= 
&\!\!\!
{{1\over 2}\left( 1 - \sum_{j=1}^N{\cal M}_j^{*2}\right) \over 
 2 {\partial T \over \partial z}}
= 
{1 - {\sum_{j=1}^N f_j^2 \over \left(\sum_{k=1}^N f_k\right)^2} 
\over 2\left(1 - {|\sum_{l=1}^N {\rm e}^{i\alpha_l} f_l|^2 \over 
\left(\sum_{m=1}^N f_m\right)^2 }\right) }
\nonumber \\
&\!\!\!
=
&\!\!\!
{1 -{\sum_{j=1}^N f_j^2 \over \left(\sum_{k=1}^N f_k\right)^2} 
\over 2\left( 1 - |T|^2 \right) } 
=
{2{\sum_{j<l} f_j f_k \over \left(\sum_{k=1}^N f_k\right)^2} 
\over 2\left( 1 - |T|^2 \right) } 
. 
\label{eq:kahler_ZN_LSM}
\end{eqnarray}
Here we have expressed the metric in terms of 
the auxiliary quantities $f_j$ as an intermediate step. 
The numerator of the right-hand side can be expressed in 
terms of the $T=T_R + iT_I$, since $f_j$ are given in terms of $z$ which 
can be expressed in terms of $T_R$ and $T_I$. 
Therefore we have solved the ${1 \over 4}$ BPS equations and obtained 
the K\"ahler metric of the nonlinear sigma model implicitly through 
Eqs.(\ref{eq:fj_general}), (\ref{eq:junction_ZN_LSM}), 
and (\ref{eq:kahler_ZN_LSM}).
It is gratifying to find that the resulting K\"ahler metric is 
real. 
This is a nontrivial requirement which should be imposed on the 
K\"ahler metric. 
The asymptotic behavior of the solution precisely reproduces 
the expected vacuum configuration and the connecting wall 
configurations. 
%the same as in the previous section. 
Therefore we have found the ${1 \over 4}$ BPS junction solution connecting 
$N$ discrete vacua. 

\begin{figure}[t]
%\begin{figure}[thbp]
\begin{center}
%\begin{flushleft}
%%%%%%%%%%%%%%%%epsfig.sty%%%%%%%%%%%
\leavevmode
\begin{eqnarray*}
\begin{array}{cc}
  \epsfxsize=4cm
  \epsfysize=4cm
\epsfbox{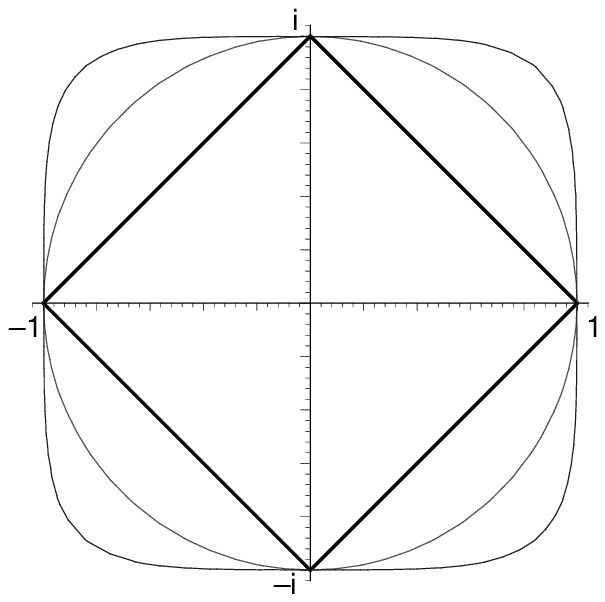} & 
%\vspace*{5cm} & 
  \epsfxsize=4cm
  \epsfysize=4cm
\hspace*{1cm}
\epsfbox{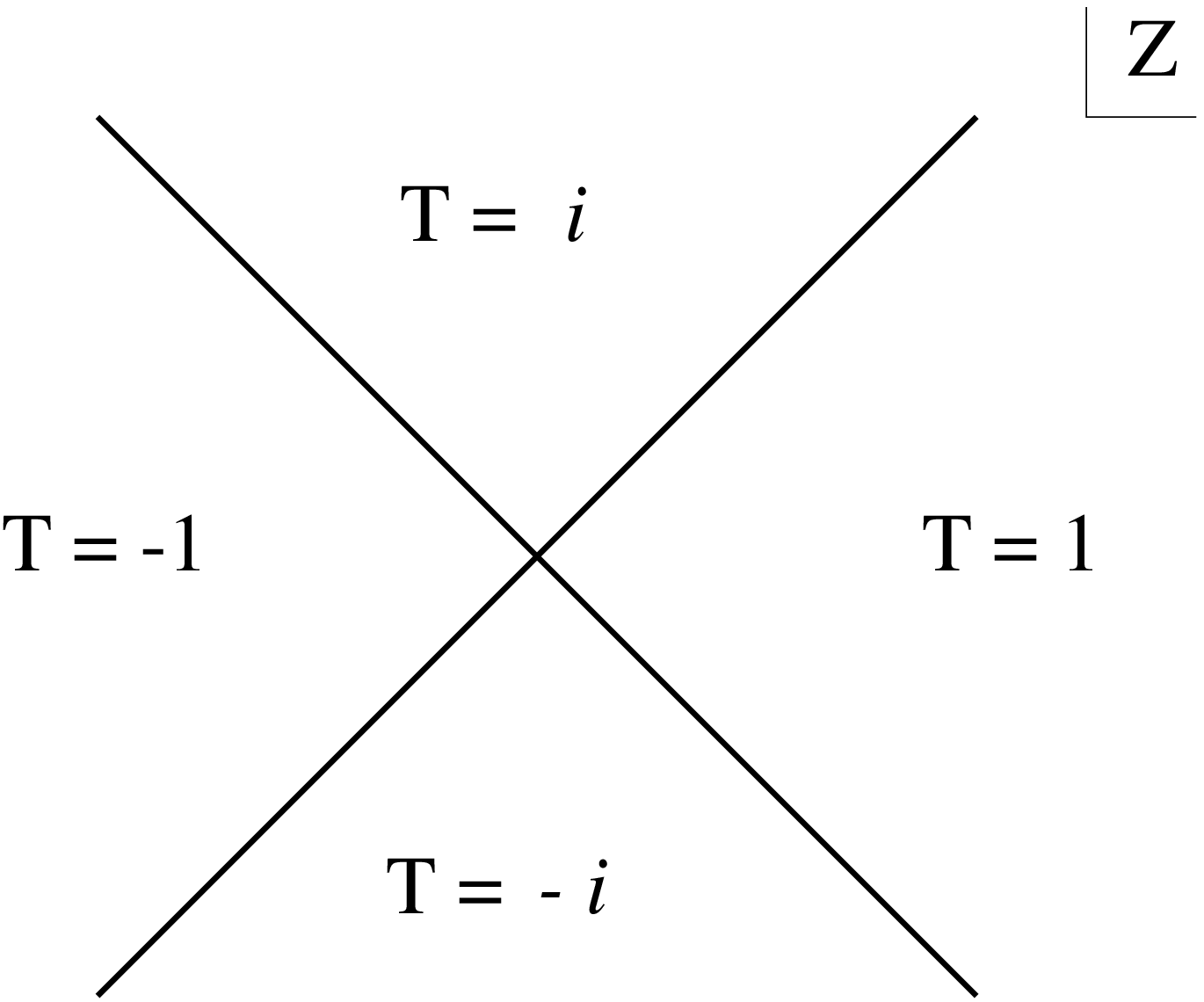} 
\\
\mbox{\footnotesize (a) Vacua of the $Z_4$ symmetric nonlinear sigma model} & 
\hspace*{1cm}
\mbox{\footnotesize (b) $Z_4$ junction}
\end{array} 
\end{eqnarray*} 
%\centerline{\epsfbox{well.ps}}
%\vspace{5cm}
%%%%%%%%%%%%%%%graphics.sty%%%%%%%%%
%%\includegraphics{well.ps}
%%%%%%%%%%%%%%%%%%%%%%%%%%%%%%%%%%%
\caption{Nonlinear sigma model with $Z_4$ symmetric junction. }
\label{MODULI-8}
\end{center}
%\end{flushleft}
\end{figure}

We now examine the properties of the K\"ahler metric 
(\ref{eq:kahler_ZN_LSM}). 
The singularity can occur only on a circle $|T|=1$. 
However, the junction solution can never touch the circle as long as 
$z$ is finite. 
The circle can be reached only asymptotically as $|z|\rightarrow \infty$. 
In fact, %the analysis in the previous section shows that 
the junction solution approaches asymptotically along a generic direction 
to one of the discrete vacua, say ${\rm e}^{i\alpha_j}$. 
On the other hand, the numerator is such that it also vanishes precisely 
at these vacua and 
the K\"ahler metric becomes finite at the discrete vacua 
\begin{equation}
K_{TT^*} \rightarrow 
{1 \over 2\left(1-\cos(\alpha_j-\alpha_{j-1})\right)}, 
\label{eq:kahler_vacuum}
\end{equation}
if the nearest vacuum to ${\rm e}^{\alpha_j}$ is at 
 ${\rm e}^{\alpha_{j-1}}$. 
If the nearest vacuum is at  ${\rm e}^{\alpha_{j+1}}$ instead, 
we should replace  $\alpha_{j-1}$ by $\alpha_{j+1}$. 
This result shows that the K\"ahler metric is in fact nonsigular as 
we have assumed in analyzing the SUSY vacua. 
The junction  asymptotically along the wall direction 
becomes a wall solution. 
In our solution, 
%we have already shown in the previous section that 
the wall is mapped to a straight line segments 
connecting adjacent vacua in the $T$ plane. 
The K\"ahler metric takes the value (\ref{eq:kahler_vacuum}) 
%$K_{TT^*}=1/(2\cos(\alpha_j-\alpha_{j-1}))$ 
along the straight line corresponding to the wall connecting the vacua 
$T={\rm e}^{\alpha_{j-1}}$ and $T={\rm e}^{\alpha_{j}}$. 
Let us call $R=\{T(z, z^*) \in {\bf C}, z \in {\bf C}\}$ 
the image of the entire real space $z$ by the map $T(z, z^*)$ 
defined by the junction solution (\ref{eq:junction_ZN_LSM}). 
Summarizing the properties of the K\"ahler metric, we find 
\begin{enumerate}
\item 
The K\"ahler metric is real and positive in $R$. 
\item 
The K\"ahler metric is never singular in $R$. 
\item
The field $T$ can approach asymptotically to the unit circle 
only at the discrete vacuum $T={\rm e}^{i\alpha_j}$ where 
the K\"ahler metric is finite. 
\item
The asymptotic value of the junction solution is mapped to a straight 
line segments connecting these discrete vacua in the $T$ plane. 
\item
At the origin of the base space, the auxiliary quantity becomes $f_j=1$. 
In the case of $Z_N$ symmetric model, 
the field takes values $T=0, {\cal M}_j=1/N$ and 
the K\"ahler metric takes a value $K_{TT^*} = (N-1)/(2N)$. 
\end{enumerate}
These results suggest that $R$ is a polygon with the $N$ discrete vacua 
as vertices. 
It is interesting to observe that the moduli space is divided into 
several discrete stationary points and 
singular arcs as illustrated by an example in Fig.\ref{MODULI-8}. 
We expect 
that the kinetic term is positive 
definite in $|T| \le 1$, and that the natural domain 
of definition for our nonlinear sigma model is $|T|\le 1$, 
which turn out to be the case in models that we are going to 
analyze more explicitly. 

Let us evaluate the K\"ahler metric as a function of $T$ more explicitly. 
{}For that purpose, we shall first take the $Z_N$ symmetric case. 
The model with junction starts from $N=3$. 
The model with $N=3$ turn out to give a minimal kinetic term 
$K_{TT^*}=1$ and hence 
reduces to our original model in Ref.\cite{OINS}. 
The next model is $N=4$. 
We find that the model in fact gives a nonlinear sigma model 
for 
the field $T$ as 
\begin{equation}
K_{TT^*} = { 1 \over 2(1 - |T|^2)} 
\left[{3 \over 4} -{1 \over 2}|T|^2 
-{1 \over 16}\left(T^4 
+T^{*4}\right) 
-{1 \over 8} |T|^4 \right] .
\label{eq:kahler_ZN_T}
\end{equation}

\begin{figure}[htbp]
\begin{center}
%%%%%%%%%%%%%%%epsfig.sty%%%%%%%%%%%
\leavevmode
\epsfxsize=8cm
\epsfysize=5cm
\centerline{\epsfbox{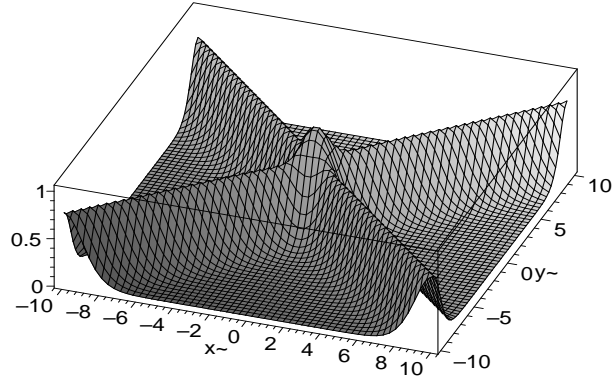}}  
\caption{Energy density of $Z_4$ symmetric junction in $x^1, x^2$ plane 
with vacua at $1, i, -1, -i$. 
}
\label{FIG:Z4symmetric_junction}
\end{center}
\end{figure}

%%%%%%%%%%%%%%%%%%
As shown in Fig.\ref{MODULI-8}, 
there is a locus of zeros of the numerator of the K\"ahler metric 
which touches the unit circle precisely 
at the four discrete points $T={\rm e}^{i{2\pi \over 4}j}, j=1, \cdots, 4$. 
This concrete example shows our general analysis clearly: 
 the K\"ahler metric is singular along 
four arcs separated by discrete stationary points of the superpotential 
$T={\rm e}^{i{2\pi \over 4}j}, j=1, \cdots, 4$ 
where the K\"ahler metric is finite. 
The region between the circle $|T|=1$ and the outer curve touching the circle 
at $T=1, i, -1, -i$ gives the region of negative kinetic term 
as shown in Fig.~\ref{MODULI-8}(a). 
The energy density of the junction is shown in 
 Fig.~\ref{FIG:Z4symmetric_junction}.  
In this model, we have also a continuous moduli space consisting of 
the arcs of singularities of the K\"ahler metric. 
However, even if we attempt to make a wall and/or a junction solution 
connecting these vacua of singularities, we find that 
the BPS equations are not satisfied by Ansatz similar to 
 our previous one in Eqs.(\ref{eq:junctionAnsatz}) or 
(\ref{eq:junction_ZN_LSM}), since 
$T$ now has a nonlinear kinetic term in our model. 
It is possible that there may exist wall or junction solutions 
connecting these singularities of the K\"ahler metric. 
Even if they exist, however, these solutions cannot be adiabatically 
deformed to our solution, since the other fields ${\cal M}$ take 
different values in two types of vacua as shown in Eqs.(\ref{eq:ZNvacLSM})
and (\ref{eq:kahlersing}). 

As another example, we shall give a model without the $Z_N$ symmetry. 
Let us take three ``matter'' fields ${\cal M}_j, j=1,2,3$ besides $T$. 
We choose the superpotential (\ref{eq:N_superpotential}) 
with $\alpha_1=\pi/2, \alpha_2=-\pi/2, \alpha_3=0$. 
Then we have 
three discrete vacua at asymmetric points $T=1, i, -i$ on the unit 
circle. %as shown in Fig.~\ref{FIG:asymmetric_junction}. 
\begin{figure}[bhtp]
\begin{center}
%%%%%%%%%%%%%%%epsfig.sty%%%%%%%%%%%
\leavevmode
\epsfxsize=8cm
\epsfysize=5cm
\centerline{\epsfbox{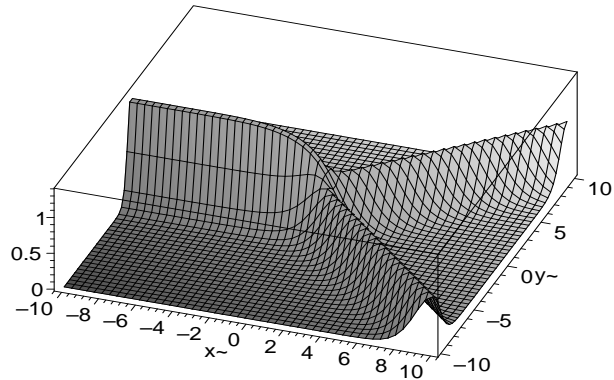}}  
\caption{Energy density of asymmetric junction in $x^1, x^2$ plane 
with vacua at $1, i, -i$. 
}
\label{FIG:asymmetric_junction}
\end{center}
\end{figure}

We find that the K\"ahler metric in Eq.(\ref{eq:kahler_ZN_LSM}) 
 becomes in this model 
\begin{equation}
K_{TT^*} = { 1 \over 4(1 - |T|^2)} 
\left(1 +T+T^* - 2|T|^2 
-{T^2+T^{*2} \over 2}\right). 
\label{eq:kahler_asym_T}
\end{equation}
The K\"ahler metric again has a general feature: 
it is finite at the discrete vacua which separate the arcs of singular 
points of the K\"ahler metric. 
%as shown in Fig.~\ref{FIG:asymmetric_junction}. 
The kinetic term is negative in the region between 
the circle $|T|=1$ and the ellipse touching the circle 
at three vacua $T=1, i, -i$. 
The energy density of the junction is shown in 
 Fig.~\ref{FIG:asymmetric_junction}.

%\section{Conclusions}

%\noindent
%\textbf{Acknowledgments.}  One of the authors (N.S) thanks a discussion}.

\end{document}